\documentclass[11pt]{article}
\usepackage{amsmath,amsfonts}
\allowdisplaybreaks
\def \beq  {\begin{equation}}
\def \eeq  {\end{equation}}
\def \beqar {\begin{eqnarray}}
\def \eeqar {\end{eqnarray}}

\def\be{\begin{equation}}
\def\ee{\end{equation}}
\def\bsp{\be\begin{split}}
\def\sqr#1#2{{\vcenter{\vbox{\hrule height.#2pt
\hbox{\vrule width.#2pt height#1pt \kern#1pt
\vrule width.#2pt}\hrule height.#2pt}}}}

\def\la {{\langle}}
\def\ra {{\rangle}}

\def\vf {{\varphi}}

\def\Tr {{\rm Tr}}
\def\tr{\mbox{tr}}

\def\bu {\bar{u}}

\def\vf {{\varphi}}

\def\del {\partial}

\def\g{\gamma}

\def\e {\epsilon}

\def\m{\mu}

\def\s{\sigma}

\def\H {{\cal H}}

\def\half{\textstyle{1\over 2}}

\begin{document}
\fontfamily{cmr}\fontsize{12pt}{15pt}\selectfont
\def \CMP {{Commun. Math. Phys.}}
\def \PRL {{Phys. Rev. Lett.}}
\def \PL {{Phys. Lett.}}
\def \NPBProc {{Nucl. Phys. B (Proc. Suppl.)}}
\def \NP {{Nucl. Phys.}}
\def \RMP {{Rev. Mod. Phys.}}
\def \JGP {{J. Geom. Phys.}}
\def \CQG {{Class. Quant. Grav.}}
\def \MPL {{Mod. Phys. Lett.}}
\def \IJMP {{ Int. J. Mod. Phys.}}
\def \JHEP {{JHEP}}
\def \PR {{Phys. Rev.}}
\def \JMP {{J. Math. Phys.}}
\def \GRG{{Gen. Rel. Grav.}}
\begin{titlepage}
\null\vspace{-62pt} \pagestyle{empty}
\begin{center}
\rightline{CCNY-HEP-15/2}
\rightline{April 2015}
\vspace{1truein} {\Large\bfseries
Fermions, Mass-Gap and Landau Levels:}\\
\vskip .1in
{\Large\bfseries
Gauge invariant Hamiltonian for QCD in D=2+1}\\
\vspace{6pt}
\vskip .1in
{\Large \bfseries  ~}\\
\vskip .1in
{\Large\bfseries ~}\\
{\large\sc Abhishek Agarwal$^{a,b}$ and V.P. Nair$^a$}\\
\vskip .2in
{\itshape $^a$\,Physics Department\\
City College of the CUNY\\
New York, NY 10031}\\
\vskip .1in{\itshape $^b$\,Physical Review Letters\\
American Physical Society\\ 
Ridge, NY 11961}\\
\vskip .1in
\begin{tabular}{r l}
E-mail:
&{\fontfamily{cmtt}\fontsize{11pt}{15pt}\selectfont abhishek@aps.org}\\
&{\fontfamily{cmtt}\fontsize{11pt}{15pt}\selectfont vpn@sci.ccny.cuny.edu}
\end{tabular}

\fontfamily{cmr}\fontsize{11pt}{15pt}\selectfont
\vspace{.8in}

\centerline{\large\bf Abstract}
\end{center}
A gauge-invariant reformulation of QCD in three spacetime dimensions is presented within a Hamiltonian formalism, extending previous work to include fermion fields in
the adjoint and fundamental representations. 
{\it A priori} there are several ways to define the gauge-invariant versions of the fermions;
a consistent prescription for choosing the fermionic variables is presented. 
The fermionic contribution to the volume element of the gauge orbit space and the gluonic mass-gap is computed exactly and this contribution is shown to be closely related to the mechanism for induction of Chern-Simons terms by parity-odd fermions. The consistency of the Hamiltonian scheme with known results on index theorems, Landau Levels and renormalization of Chern-Simons level numbers is shown in detail. We also comment on the fermionic contribution to the volume element
in relation to issues of confinement and screening.
\end{titlepage}
\pagestyle{plain} \setcounter{page}{2}
\fontfamily{cmr}\fontsize{12pt}{15.5pt}\selectfont
\section{Introduction and Summary}
In this paper we extend the gauge invariant Hamiltonian formulation of gauge theories
in two spatial dimensions, developed in \cite{KKN1} to include quarks in both fundamental and adjoint representations. In particular, we compute the quark contribution to the volume of the gauge invariant configuration space of $QCD_{2+1}$. The volume element for the gauge orbit space (configuration space) of gauge theories, which is also the measure of functional integration for physical field configurations,
 is an important quantity which has direct implications for the existence of a mass gap 
 and for other questions of physical interest. The results obtained in this paper allow us to 
 analyze how quarks affect the mass-gap of the theory; the answer, we find,
  depends  both on the representation of $SU(N)$ carried by the quarks, as well as on their parity properties. Even though we work with non-supersymmetric theories in this paper, our answers are in complete agreement with expectations from the supersymmetric models studied in \cite{AA-VPN}. 
The formal gauge invariant reformulation of $QCD_{2+1}$ and
fermionic contribution to the mass gap are
 exact and nonperturbative. And in the limit of the quarks being veryheavy, we do recover 
 the confinement and screening properties as expected from the representation and parity 
 properties of the quarks.

To further motivate the analysis of this paper, it is useful to recall some salient aspects of the KKN approach to pure gluodynamics \cite{KKN1}. The starting point there was the parameterization of the spatial components of the gauge potentials $A = \frac{1}{2}(A_1 + iA_2)$ and $\bar A = \frac{1}{2}(A_1 - iA_2)$ in terms of the $SL(N, \mathbb{C})$-valued complex matrices $M$ and $M^\dagger$ as\footnote{$\partial = \frac{1}{2}(\partial_1 + i\partial_2)$ and $\bar \partial = \frac{1}{2}(\partial_1 - i\partial_2)$, and $A_0 = 0$ as is appropriate for a Hamiltonian set up.}
\be
A = -\partial MM^{-1}, \hspace{.3cm} \bar A = M^{\dagger -1}\bar \partial M^{\dagger}\label{M}
\ee
Since $M \rightarrow UM$ under local time independent gauge transformations \cite{KKN1}, the above parameterization allows one to readily identify 
$H = M^\dagger M\label{H}$
as the natural gauge invariant variable for pure Yang-Mills in $D = 2+1$.  
We may regard $H  \in SL(N, \mathbb{C})/ SU(N)$ as furnishing a coordinatization of the space of gauge-invariant configurations
$\mathcal{C}$, in the pure glue case.
The volume element for $\mathcal{C}$ is then given by \cite{KKN1} 
\be
d\mu[\mathcal{C}] = d\mu[H] ~e^{2 c_A S_{WZW}[H]}\label{measureH}
\ee
where $d\mu [H]$ is the Haar measure on the space of hermitian matrices
$H$
and
\be
S_{WZW}[H] = \frac{1}{2\pi}\int \tr(\partial H \bar \partial H^{-1}) +\frac{i}{12\pi}\int \epsilon^{\alpha \beta \gamma} \tr(H^{-1}\partial_\alpha HH^{-1}\partial_\beta HH^{-1}\partial_\gamma H) 
\ee
is the Wess-Zumino-Witten (WZW) action for the field $H$.
Here $c_A$ is the adjoint Casimir of the group, equal to $N$ for $SU(N)$. The $S_{WZW}$ factor
arises from the Jacobian for change of variables from $A,\bar A$ to $H$. 
The inner product for wave functions is then given by
\be
\langle \Phi_1|\Phi_2\rangle = \int d\mu[H] ~e^{2\,c_A S_{WZW}[H]}~ \Phi_1^* \, \Phi_2
\ee
Consequently, $S_{WZW}$ plays a key role in 
the self-adjointness properties of observables. It also provides a cut-off on the long 
wavelength modes of the fields with the result that
 the fundamental mass gap parameter $\Delta$ of the gluonic theory
is related to the pre-factor $\tilde k = 2c_A$ of $S_{WZW}$ as
 \cite{KKN1}
\be
\Delta = \frac{e^2 \tilde k}{4\pi}\label{gap}
\ee
Independent arguments based on self-adjointness and Lorentz invariance can be invoked to make the case that $\Delta$ and $ \tilde k$ must be related as in
 (\ref{gap}) for reasons of internal consistency \cite{robust}.

With these basic features of the pure Yang-Mills theory  in mind,
we now turn to the question of defining the gauge-invariant 
configuration space and its volume measure for Grassman-valued fermionic 
fields in a Hamiltonian 
framework.
As will be shown later, there are inequivalent prescriptions for defining
 gauge-invariant fermionic variables
 which differ in renormalizing $\tilde k$ by finite, but different amounts. 
 Since $\tilde k$ is related to the mass-gap for gluons, the choice of which prescription to use
 cannot be made in an {\it ad hoc} manner;
a consistent principle for determining the gauge invariant form of fermionic matter is needed.
Given that that $\tilde k$ and $\Delta $ are related as in (\ref{gap}),  the chosen
prescription (and the corresponding 
renormalization of $\tilde k$) must be reconciled with the self-adjointness of the Hamiltonian 
in terms of gauge-invariant variables.
This is one of the key issues addressed in this paper.

The present article builds on \cite{AA-VPN}, where we considered the contribution to the mass gap due to adjoint Majorana fermions
in $1\leq \mathcal{N}\leq 4$ supersymmetric gauge theories. It was noted there that
there are different ways of defining gauge-invariant versions of the fermion fields, and 
two specific cases of defining the gauge-invariant variables were considered in some detail.
One of them led to a Jacobian which increased $\tilde k$ by $\frac{1}{2}\tilde k$,
while the other decreased it by the same amount. Only these two choices were compatible with the Majorana condition.
Interestingly, it was found that supersymmetry helped to choose between the two possibilities.
For the ${\cal N} =1$ supersymmetric theory, the correct  choice - dictated by the requirement of manifest supersymmetry -  was the one
for which the fermionic contribution decreased the mass
gap (specifically, $\tilde k$ was decreased by a factor of half).The same choice applied, for the same reasons, to
the ${\cal N }= 2$ case as well, which meant, because of the double contribution, that
the mass gap was effectively reduced to zero.
If the tendency continued for higher extensions of supersymmetry, there would be a problem
as a negative mass gap would not be tenable.
However, requirement of manifest supersymmetry once again helps to make things consistent; this
was most vividly illustrated by the case of ${\cal N} =4$ supersymmetric theory.
In this case, three of the fermion fields (related to each other by the manifest $R$-symmetry)
contribute as before while the remaining one
had the opposite contribution, so that the same situation as in the ${\cal N } =2$ case was obtained
for the volume measure. The theories considered in\cite{AA-VPN} also included a Chern-Simons term with level number $k$. For the purely gluonic theory, the coefficient of $S_{WZW}$ is given by $\tilde k + k$, which can be identified with the renormalized Chern-Simons coupling of the $\mathcal{N} = 0$ theory. 
The different renormalizations of $\tilde k +k$ by the different choices of gauge-invariant variables
(as dictated by supersymmetry)
were shown to match with the renormalization of $k$ obtained using covariant perturbation theory \cite{YMCS-lee}.  
    
It is quite striking (and gratifying) that supersymmetry helps to select the right set of gauge-invariant variables and also that the renormalization of $k$ as given by covariant perturbation theory
can be related to an intrinsic geometric property of the full configuration space
$\mathcal{C}$ including the fermions
\footnote{$k \rightarrow k + \tilde k$ is also seen in pure Chern-Simons theory on the plane \cite{bos-nair, wittencs, rob-rao}where there are no dynamical degrees of freedom, which is a strong hint that this renormalization is a property of the
configuration space $\mathcal{C}$.}. Nevertheless, it was not clear from \cite{AA-VPN} how 
one might extend these considerations to non-supersymmetric theories, 
including Dirac fermions, fermions in the fundamental representation, etc.. These issues
are the main focus of this article.

This paper and its main results are organized as follows.
In the next section we review the connection between the measure on $\mathcal{C}$ and the mass-gap in pure gluodynamics in the KKN formalism. 
After introducing fermions,
we discuss the possible gauge-invariant choices for the fermion fields.
(which can depend on the representation of the fermions).
 We then compute the Jacobians and their contributions to $d\mu[\mathcal{C}]$ for each choice.
 In this section we outline the argument for
the compatibility of the nonzero fermionic contribution to $d\mu[\mathcal{C}]$ 
and requirements of self-adjointness. 

Following this we explore the consistency conditions that dictate the correct choice of gauge invariant variables for fundamental and adjoint fermions. For a single fermionic species, the Jacobian in transforming to gauge-invariant fermionic variables shifts the coefficient of $S_{WZW}$ 
from $\tilde k $ to $ \tilde k + k'$, with the value of $k'$ depending on the 
gauge-invariant form for the fermion field.
From covariant perturbation theory, we know that
a fermion field has a parity violating mass term $m$ will induce a Chern-Simons term in the effective action with level number $k = \frac{m}{|m|}c_R$, where $c_R$ is the index
of the  representation $R$ carried by the fermion field. This induced Chern-Simons term will, in turn, affect the dynamics of the gluons. As argued in \cite{AA-VPN} via the use of the
Knizhnik-Zamolodchikov equation, this result implies that 
the coefficient of $S_{WZW}$ in the gluonic measure would be changed to $2k + \tilde k$. 
Clearly one must require 
\be
k' = 2k\label{consistency}
\ee
for compatibility with the Jacobian calculation.
We then find that there is a $unique $ choice of gauge-invariant fermionic variables
for both fundamental and adjoint representations. 
Put differently, we find that $S_{WZW}$ arising from the
fermionic Jacobian must be viewed as the `boundary' part of a `bulk'
completion which is the induced Chern-Simons term obtained by integrating the fermions out
in covariant perturbation theory.
This is the basic principle for identifying the correct fermionic variables, at least in the cases where the fermions have parity-odd mass terms. 
Working within the Hamiltonian set-up of this paper, we obtain the induced Chern-Simons 
contribution by analyzing the lowest Landau levels for fermions in the adjoint and fundamental representations. 
The Chern-Simons contribution can be obtained from the $v.e.v$ of the charge density in a constant magnetic field. Compatibility with constraints that follow from various index theorems is also discussed.
(This paper, in fact, may be viewed as completing the arguments in \cite{AA-VPN}
by starting with the Hamiltonian framework and relating the measure to the covariant calculation,
thus complementing the argument presented in \cite{AA-VPN} based on the Knizhnik-Zamolodchikov equation.)

In the case of fermions with parity-preserving mass terms, there are
two viable options for their gauge-invariant forms. Although a unique choice does not seem to be picked out by
consistency reasons, we point out that
one of these choices is particularly well suited for computational purposes.

These basic results have implications for
how the fermionic matter content affects confinement and screening properties of $QCD_{2+1}$. This is discussed in the final section. 

\section{Gauge-invariant fermionic variables and measure}\label{jac-com}

How does the  measure of integration on the Hilbert space of a quantum system carry information about its spectrum, specifically the issue of a mass-gap? It is instructive to pose this question  for the harmonic oscillator Hamiltonian, $H_{H.O} = \frac{1}{2}p^2 + \frac{\Delta^{'2}}{2}x^2$, with mass-gap; $\Delta^{'}$. There are two ways to view the Hilbert space of $H_{H.O}.$ One could either regard the Gaussian ground state $\langle x|0\rangle  = e^{-\frac{1}{2}\Delta^{'} x^2}$ as the position space realization of the vacuum, or one could regard it as defining the measure on the Hilbert space of the theory, which can be taken to be spanned by polynomials in $x$ without the Gaussian factor.  In the second point of view, we note that the finite volume of the Hilbert space is indeed tied to the issue of mass-gap in the spectrum, as the coefficient of the Gaussian $\Delta $ is nothing but the fundamental mass-gap of the Hamiltonian.  This intuition carries over to gluodynamics in the KKN framework \cite{KKN1}. We briefly review this  the connection between the mass-gap, the volume of the configuration space and the self-adjointness of the Hamiltonian
for pure Yang-Mills theory before turning to fermions.

The gluonic Hamiltonian 
\be
\H_g = -\frac{e^2}{2}\int \frac{\delta ^2}{\delta A^a \delta \bar A^a} + \frac{1}{e^2}\int B^aB^a
\ee
when re-expressed using  gauge-invariant variables this becomes \cite{KKN1}
\be
 \H_g  = +\frac{e^2}{2}\int H^{ab}(x)(\bar{\cal{G}}\bar{p}^a)(x){\cal{G}}p^b(x)
+ \frac{2\pi^2}{e^2c_A^2}\int(\bar \partial J^a \bar \partial J^a)\label{KKN}
\ee  
Here, it is understood that
\begin{eqnarray}
{\cal{G}}(x,y) = G(x,y)\left[ 1 - e^{-|x-y|^2/\e}H^{-1}(y,\bar x)H(y,\bar y)\right]\nonumber \\
{\bar{\cal{G}}}(x,y)  = \bar{G}(x,y)\left[ 1 - e^{-|x-y|^2/\e}H(x,\bar y)H^{-1}(y,\bar y)\right]
\end{eqnarray}
are the point-split-regularized version of the two-dimensional Green's functions $\bar{G}(x,y) = \frac{1}{\pi(x-y)}$ and  $G(x,y) = \frac{1}{\pi(\bar x-\bar y)}$, while  ${\cal{G}}p(x)$ is a shorthand for $\int_u {\cal{G}}(x,u)p(u)$ and ${\bar{\cal{G}}}\bar p(x) = \int_u {\bar{\cal{G}}}(x,u)\bar p(u)$. $p$ and $\bar p$ are the right and left translation operators on $M$ and $M^\dagger$, acting as
\begin{eqnarray}
[p^a(y), M(x)] = M(x)(-it^a)\delta^2(x-y),\hskip .2in  [\bar{p}^a(y), M^\dagger(x)] = -it^aM^\dagger(y)\delta^2(x-y)
\end{eqnarray}
The current $J$ is gauge-invariant as well, being related to $H$ as $J = \frac{c_A}{\pi}\partial H H^{-1}$. 

The physical configuration space is the set of all hermitian matrices $H$
and, as shown in\cite{KKN1}, the measure on the space of spatial gauge potentials 
$d\mu[\mathcal{A}]$ is related to $d\mu[H]$ as
\be
d\mu[\mathcal{A}] = \det(D\bar D)\, d\mu[H]\, d\mu[\mathcal{U}]
= d\mu [H]\, e^{2 \, c_A \, S_{WZW}(H)} ~ d\mu [ {\cal U}]
\ee
The two-dimensional Dirac determinant, $\det(D\bar D)$, is the Jacobian for the change of variables. and we have used the fact that, up to unimportant constants, it is given by
$e^{2c_A S_{WZW}[H]}$. $\mathcal{U}$ denotes the space of gauge transformations. Since the 
gauge-invariant configurations space is $\mathcal{A}/\mathcal{U}$, $d\mu[\mathcal{C}]  = \det(D\bar D)d\mu[H]$ which leads us back to (\ref{measureH}). Comparing with the harmonic oscillator, it is evident that $e^{2c_AS_{WZW}}$ is the analog of the gaussian, with the prefactor $2\,c_A$  playing the 
role of $\Delta'$. 
This is further supported by
 by looking at the action of (\ref{KKN}) on single $J$ fields\cite{KKN1} (which is the gauge theory  analog of acting with $H_{H.O}$ on $a^\dagger |0\rangle$).  Finally, we also recall that the coefficient of $S_{WZW}$ needs to be what it is on grounds of self-adjointness. The self-adjoint nature of the kinetic energy operator requires that
\be
(\bar{\cal{G}}\bar{p}^a)(x) = \left[{\cal{G}}p^a(x)\right]^\dagger\label{adjoint}
\ee
This relation can be established using the identity
\be
[\bar{\cal{G}}\bar{p}^a(x), H^{ab}(x)e^{2\,c_AS_{WZW}[H]}] = 0
\ee
which is obtained only if the coefficient of $S_{WZW}$ is $2\,c_A$.

We now turn to fermions and outline
the possible ways of defining gauge-invariant variables and the Jacobian for each case for the corresponding change of variables.
Following the conventions of \cite{AA-VPN} and take the three dimensional gamma matrices to be $\{\gamma^\m = i\s^3, \s^1, \s^2\}$. A general Dirac spinor has two components which we may 
label $\psi_1, \, \psi_2$, defined by
\be
\Psi = \left(\begin{array}{c}
\psi_1\\
\psi_2
\end{array}\right)\label{1}
\ee
The charge conjugation matrix $C$ defined by
$ C\gamma ^\mu C^{-1} = -(\gamma ^\mu)^t$ (where the superscript $t$ denotes the transpose)
can be taken to be $\s^2$. When the fermions are in the adjoint representation we can impose  the Majorana condition $\bar \Psi = \psi ^t C$; which  amounts to the identification  $\psi_2 = \psi_1^*$.  In the Hamiltonian framework, we are interesting in the volume measure for two-dimensional fields since this is what is relevant for the integral for
the inner product of wave functions.

Focussing for the moment on adjoint Dirac fermions, we see that there are four available choices for generating gauge invariant fermionic variables. Denoting the gauge-invariant two-component fermions by $\Lambda $ (and the components by $\chi_i$), the choices are: \vskip .1in\noindent
{\bf Choice I}\\
\be
\Lambda = \left(\begin{array}{c}\chi_1\\ \chi_2 \end{array}\right)  = \left(\begin{array}{c}
M^{-1}\psi_1\\ M^{\dagger } \psi_2\end{array}\right) 
\ee
{\bf Choice II}\\
\be
\Lambda = \left(\begin{array}{c}\chi_1\\ \chi_2 \end{array}\right)  = \left(\begin{array}{c}
M^{\dagger } \psi_1\\ M^{-1}\psi_2\end{array}\right) 
\ee
{\bf Choice III}\\
\be
\Lambda = \left(\begin{array}{c}\chi_1\\ \chi_2 \end{array}\right)  = \left(\begin{array}{c}
M^{\dagger } \psi_1\\ M^{\dagger}\psi_2\end{array}\right) 
\ee
{\bf Choice IV}\\
\be
\Lambda = \left(\begin{array}{c}\chi_1\\ \chi_2 \end{array}\right)  = \left(\begin{array}{c}
M^{-1 } \psi_1\\ M^{-1}\psi_2\end{array}\right) 
\ee 
It is understood that $M^{ab} = 2\,\tr (t^aMt^bM^{-1})$ is the adjoint representative of the matrix $M$ in the above choices.
Of the four choices, only I and II are compatible with the Majorana condition, while all four are, in principle, valid choices for Dirac fermions.  If we change variables to the $\chi$'s
according to these choices, there will be nontrivial Jacobians for the measure for the
fermionic configuration space.
To analyze these Jacobians, we will first need to define the measure which is best done
using fermionic coherent states.

For a Dirac spinor $\Psi$ with components $\psi_1, \, \psi_2$, we can introduce
independent Grassmann variables $U = (u_1, u_2)$, $\bar U = (\bu_1, \bu_2)$and define 
\be
|U\rangle = e^{-\bar U \Psi}|0\rangle, \hspace{.2cm} \langle U| = \langle 0| e^{-\check \Psi U}
\label{6}
\ee
By using standard properties of coherent states, it can be shown that $\Psi_i$ is represented by the multiplication operation $u_i$ on the coherent states while, its canonical conjugate variable $\check \Psi_i \equiv \frac{\delta}{\delta u_i}$. In the coherent state picture: $\check \Psi_i \equiv \frac{\delta}{\delta u_i} \equiv \bar u_i$. In fact, any normal-ordered operator $\hat A(\check \Psi,  \Psi)$ can be replaced by its symbol, which is the Grassmann valued function $A(\bar U, U)$ while looking at coherent state expectation values.  Furthermore, the symbol of products of normal-ordered 
operators $\hat A, \hat B$ is given by $A \star B$, where $\star$ is the Grasssmann version of the Wigner-Moyal star product\footnote{For previous work on fermions in a functional Scrodinger representation see \cite{jackiw, wipf, mansfield}.}. Thus, in a Hamiltonian picture,  wave functions are functions of $u_i$ and the norm and overlap of wave functions are computed using the functional measure
\be
\mathcal{D} U \mathcal{D}\bar U \,e^{-\int U\bar U} \label{7}
\ee
which is identical to the measure for Euclidean functional path integrals for fermions in two dimensions. 
(The Gaussian factor may be considered as $e^{-K}$ where $K$ is the K\"ahler potential
for the fermionic coherent states.)
We can now expand the Grassmann fields in a complete set of modes as
\be
U  = \sum a_n \phi_n, \hspace{1cm} \bar U = \sum b_n \phi^*_n \label{8}
\ee
where $a_n, b_n$ are independent Grassmann variables.  The measure, $d\mu$, (disregarding the Gaussian part for now) on the space of the wave functions  is given formally by
\be
d\mu = \mathcal{D}U \mathcal{D}\bar U = \Pi_n db_n \Pi_m da_m\label{9} 
\ee
The expression above can be taken to define the measure on the fermionic Hilbert space.
The fermionic variables involved here are related to the canonically conjugate variables,
so that the integration measure is the phase space measure for fermions.
We can now consider the
change in the measure under transformations of variables, giving formal definitions of the Jacobians which we will calculate more carefully in a regularized way.
Fermions which transform as the adjoint representation and under the
fundamental representation have different behaviors and will be considered separately.

\subsection{Adjoint Dirac fermions}

Writing $M = e^{\theta ^aT^a}$ (where $T^a$ are the Hermitian generators of the gauge group in the adjoint representation), the choices I and II amount, in infinitesimal form, to 
\be
\Lambda  = (I - i\,\Im \theta^a\,  T^a \mp\, \Re \theta^a\, T^a\gamma^5 + \cdots)\Psi
\label{10}
\ee
$\gamma^5 = \sigma^3$ which acts as a $\gamma^5$ in the two-dimensional sense. The change to gauge-invariant variables is thus reminiscent of a chiral transformation in two Euclidean dimensions. The Jacobian for this change of variables is thus related to
a two-dimensional chiral anomaly.
The upper and lower signs in (\ref{10}) correspond to choices I and II respectively. It is easy to see that the remaining two choices do not involve $\gamma^5$ and hence
they do not have nontrivial Jacobians associated with them.  The canonically conjugate gauge-invariant fields for I and II are
\beqar
\check \Lambda &=& \left(\begin{array}{c}\check \chi_1\\ \check \chi_2 \end{array}\right)  = \left(\begin{array}{c}
M^{-1}\psi^*_1\\ M^{\dagger } \psi^*_2\end{array}\right) \hskip .3in ({\rm Choice~I}) \nonumber\\
\check \Lambda &=& \left(\begin{array}{c}\check \chi_1\\ \check \chi_2 \end{array}\right)  = \left(\begin{array}{c}
M^{\dagger } \psi^*_1\\ M^{-1}\psi^*_2\end{array}\right) \hskip .3in ({\rm Choice ~II})
\label{11}
\eeqar
Using the fact that $M^t = M^{-1}$ in the adjoint representation, it is readily verified that $\{\check \chi _i, \chi_j\} = \delta _{ij}$
It is also understood that for the Dirac spinors $\psi^*_i = \check \psi_i$. Thus, in the infinitesimal form,
\be
\check \Lambda  = (I - i\Im \theta^a  T^a \mp \Re \theta^a T^a\gamma^5\cdots)\check \Psi
\label{12}
\ee
Writing $\Lambda  = a'_n \phi_n$, $\check \Lambda  = b'_n \phi^*_n$, from (\ref{10}, \ref{12}),
the transformation of the mode variables  is seen to be
\be
a'_n = \sum_m C_{nm} a_m, \hspace{.2cm} b'_n = \sum_mC_{mn}b_m \label{13}
\ee
where
\be
C_{nm} = \int \phi^*_n(I - i\Im \theta^a  T^a \mp \Re \theta^a T^a\gamma^5\cdots)\phi_m
\label{14}
\ee
The Jacobian for the transformation (\ref{13}) is thus given by
\be
J = \det C^2 \label{15}
\ee
As mentioned before, the
choices III and IV do not involve $\gamma^5$ in their infinitesimal forms and hence
 the corresponding Jacobians are trivial\footnote{We refer to standards textebooks, e.g\cite{an-book} for detailed discussions about evaluation of functional determinants of the type encountered here.}.
 
\subsection{Adjoint Majorana fermions} 
 
 For Majorana fermions, we have $\psi_2 = \psi_1^*$ and the mode expansion takes the form
\be
 \left(\begin{array}{c}
\psi_1\\  \psi_1^*\end{array}\right)  = \sum_n a_n\psi_n \label{16}
\ee
Integration over fermions has to be done with the measure
$ \Pi_n da_n$\footnote{ The absence of independent  $b_n$-variables in the mode expansion is seen most transparently by taking a real representation of the $\gamma$ matrices, in which case a Majorana fermion is just represented by two real Grassmannn variables that are their own canonical conjugates.}. 
Thus, the Jacobian for the Majorana case is 
\be
J = \det C \label{17}
\ee
 which is the positive square root of the Jacobian for the Dirac case.
 
The computation of the Jacobian is standard, done by 
integrating the anomaly, see Appendix A.
The result is
 \be
 \det C =  e^{\pm (c_A S_{WZW}[H])}\label{18}
 \ee
Thus, for Dirac fermions, the Jacobian is given by $e^{\pm (2c_A S_{WZW}[H])}$
with the upper and lower signs corresponding to the choices
I and II.
For Majorana fermions the corresponding result is
 $e^{\pm (c_A S_{WZW}[H])}$. The other two choices (III and IV, and applicable only to Dirac fermions) have trivial Jacobians.
 
\subsection{Fermions in the fundamental representation} 

In the fundamental representation $ M^t \neq M^{-1}$ and the expressions for the canonically conjugate gauge-invariant fermionic variables are different from what we had for the adjoint representation above. The gauge-invariant conjugate variables are defined by
\beqar
\check \Lambda &=& \left(\begin{array}{c}\check \chi_1\\ \check \chi_2 \end{array}\right)  = \left(\begin{array}{c}
\psi^*_1M\\  \psi^*_2M^{\dagger -1}\end{array}\right) \hskip .3in ({\rm Choice ~I}) \nonumber\\
\check \Lambda &=& \left(\begin{array}{c}\check \chi_1\\ \check \chi_2 \end{array}\right)  = \left(\begin{array}{c}
\psi^*_1M^{\dagger -1}\\\ \psi^*_2M\end{array}\right) \hskip .3in ({\rm Choice~ II})
\label{19}
\eeqar
In the infinitesimal form
\be
\check \Lambda_\pm  = (I + i\Im \theta^a  (T^t)^ a \pm \Re \theta^a (T^t) ^a\g^5+\cdots) \check\Psi
\ee
where $T^t$ now denote the transpose of the  fundamental generators. Thus, in the fundamental representation, up to terms of $\mathcal{O}(\theta)$, we find
\begin{eqnarray}
J = \det C \det{\tilde C}\hspace{4cm}\nonumber \\
C = (I - i\Im \theta^a  T^a \mp \Re \theta^a T^a\g^5 )\hspace{1cm}\nonumber\\
\tilde C =  (I - i\Im \theta^a  (-T^t)^ a \mp \Re \theta^a (-T^t) ^a\g^5)\hspace{.1cm}
\end{eqnarray}
For $\tilde C$, the basic difference is that $T \rightarrow -T^t$ compared to $C$.
Since this trasnformation denotes the charge conjugation in the Lie algebra, we see that 
the determinants of $\tilde C$ and $C$ will be related by charge conjugation. The 
determinants are the same, since they only involve the quadratic invariant
$\Tr\, T^a T^b = \Tr \,(-T^t)^a (-T^t)^b$.
 Thus
\be
\det C = \det \tilde C
\ee
This is  shown more explicitly in the first Appendix.

 We have thus obtained the Jacobian contributions associated with all possible methods of generating gauge invariant fermionic variables for adjoint and fundamental fermions. We find that the Jacobian is trivial for both the adjoint and fundamental representation when choices III or IV are invoked. Adjoint Majorana fermions can contribute $e^{\pm (c_A S_{WZW}[H])}$ to the measure via the Jacobian, with the upper and lower signs corresponding to choices I and II respectively.  Furthermore, I and II are the only choices compatible with the Majorana condition. The same choices result in the the contribution  $e^{\pm (2\,c_R S_{WZW}[H])}$ when one has Dirac fermions, where
 $c_R$ is defined by $\Tr ( T^a T^b )_R = c_R \, \delta^{ab}$, for representation
 $R$. Thus
 $c_R = c_A = N$ in the adjoint representation and $c_R = \half$ for $N\geq 2$ for the fundamental
 representation. For the Abelian theory $c_R = 1$. 

 These results show that a fermionic contribution can alter the measure on $\mathcal{C}$ by shifting the coefficient of $S_{WZW}$ (and hence the mass-gap in the gluonic sector) from 
 the pure-glue case. 
 Two questions arise naturally: How is $\tilde k \neq 2\,c_A$ consistent with the self-adjointness of the Hamiltonian? How does one, on {\it a priori} grounds, select the choice of
 fermionic gauge-invariant variables?
 We turn to these questions now.
 
\subsection{Self-adjointness of gluonic Hamiltonian} 

Self-adjointness issues manifest themselves in the kinetic energy operator (which is the Laplacian on the space of gauge potentials), whose gauge-invariant form is \cite{KKN1}
\be
T =  -\frac{e^2}{2}\int \frac{\delta ^2}{\delta A^a \delta \bar A^a} = +\frac{e^2}{2}\int H^{ab}(x)(\bar{\cal{G}}\bar{p}^a)(x){\cal{G}}p^b(x)
\ee
Acting on functionals of the $J$ fields alone\footnote{The $l.h.s$ is the properly regularized version of the kinetic energy operator. The expression on the $r.h.s$ is only valid so long as one acts on $J$ fields that are at non-conincident points. These and many other regularization related issues have been discussed in \cite{KKN1}.},
\be
+\frac{e^2}{2}\int H^{ab}(x)(\bar{\cal{G}}\bar{p}^a)(x){\cal{G}}p^b(x) = \frac{e^2c_A}{2\pi}\left(\int J^a\frac{\delta}{\delta J^a} + \int \Omega ^{ab}(xy)\frac{\delta}{\delta J^a(x)}\frac{\delta}{\delta J^b(y)}\right)
\ee
where
\be
\Omega^{ab}(x,y) = [\mathcal{D}_x\bar G(y,x)]^{ab}, \hspace{.1cm} \mathcal{D}_x = \frac{c_A}{\pi}\partial_x\delta^{ab} + if^{abc}J^c(x)
\ee
The expression of $T$ in terms of $J$ shows how the mass gap could arise;
the $J\frac{\delta}{\delta J}$ term assigns energy $m$ to each power of $J$.
As reviewed earlier in this section, self-adjointness of $T$ requires $\tilde k = 2\,c_A$. 

Now, since a fermionic Jacobian could change $\tilde k$, it is natural to ask what the self-adjoint form of $T$ is when the coefficient of $S_{WZW}[H]$ is $2c_A + k'$, for some as-yet-undetermined integer $k'$. Computing $({\cal{G}}p)^\dagger$ using $e^{(2c_A +k')S_{WZW}[H]}$ in the measure,
we 
obtain
\be
({\cal{G}}p^a)^\dagger = (\bar{{\cal{G}}}\bar{p})^a - \frac{ik'}{2\pi}(\del HH^{-1})^a\label{adjointn}
\ee 
Thus, in the case of an arbitrary (nonnegative) coefficient of $S_{WZW}[H]$, the gauge invariant version of the gluonic kinetic energy operator must be defined to be
\be
T = +\frac{e^2}{2}\int H^{ab}(x)\left[{\cal{G}}p^a\right]^\dagger(x){\cal{G}}p^b(x)\label{T}
\ee
On regularized functionals of $J$, the 
self-adjoint form of $T$ corresponding to  the measure factor $e^{(2c_A + k')S_{WZW}[H]}$ is thus given by
\begin{eqnarray}
+\frac{e^2}{2}\int H^{ab}(x)\left[{\cal{G}}p^a\right]^\dagger(x){\cal{G}}p^b(x) = \frac{e^2c_A}{2\pi}\left(\int J^a\frac{\delta}{\delta J^a} + \int \Omega ^{ab}(xy)\frac{\delta}{\delta J^a(x)}\frac{\delta}{\delta J^b(y)}\right)\nonumber\\
 + \frac{e^2k'}{4\pi}\left(\int J^a\frac{\delta}{\delta J^a}\right)\hspace{5cm}\label{T1}
\end{eqnarray}
The last term on the right hand side of (\ref{T}) reflects the change in the coefficient of $S_{WZW}$. Since the fundamental mass-gap for the gluonic spectrum is generated by the $\int J\frac{\delta}{\delta J}$ term in the Hamiltonain above, we clearly see that a fermionic contribution to the measure will affect the mass-gap for the gluons. Thus, (\ref{T}) provides  a natural self adjoint kinetic energy operator when $\tilde k$ is shifted by fermionic contributions. This analysis simply amounts to the statement that a consitent hamiltonian formulation requires that the  adjoints of operators to be computed only $after $ all (both gluonic and fermionic) contributions to the measure on $\mathcal{C}$ have been accounted for. 
\section{Gauge invariant variables, Chern-Simons action and parity violating fermionic mass-terms}

We still need a principle for removing the ambiguity in the selection of fermionic variables. This is the subject of  the present section.

From the analysis of the previous section,   we note that measure factor $e^{(2c_A + k')S_{WZW}[H]}$,
where $k'$ encodes the putative contribution from a fermionic Jacobians,  is reminiscent of the measure on the gluonic configuration space for the YMCS system without fermions, but  with a Chern-Simons term with level number $k_{CS}$ \cite{YMCS}. (The measure factor for the
YMCS theory is $e^{(2c_A + k_{CS})S_{WZW}[H]}$ and replacing $k'$ by $k_{CS}$ in (\ref{T}) correctly reproduces gauge invariant expression for $T$ corresponding to the YMCS system.)
The parallel between the measures for the YMCS system and that of the effective measure for the gluons with a nontrivial contribution from the fermionic Jacobian suggests that the renormalization of a bare $k_{CS}$, and the Jacobian contribution from fermions are deeply related quantities. This is indeed the case and prompts us to study the cases of fermions with parity violating and parity preserving mass terms separately.

\subsection{Parity violating fermions in the fundamental representation} 

We first focus first on $QCD_{2+1}$ with a single flavor of fermions transforming in the fundamental representation of $SU(N)$. This is obviously a parity-violating theory and there will be an induced
CS term upon integrating out the fermions. The Hamiltonian is given by
\be
\H = -\frac{e^2}{2}\int \frac{\delta ^2}{\delta A^a \delta \bar A^a}  + \frac{e^2}{2}\int B^aB^a + \frac{i}{2}\int \bar \Psi^a \left[(\gamma^i D_i - m)\Psi\right]^a\label{QCDF}
\ee
Here $D_i = \del_i - i A_i^a\,t^a$ and $t^a$ are hermitian matrices which form a basis
for the Lie algebra of $SU(N)$. The gauge and Lorentz-invariant fermionic mass term $\bar \Psi \Psi$ breaks parity invariance. With multiple flavor species of fermions, one can define alternate mass terms which are parity-invariant. However, we will focus first on a single fermionic species with a parity breaking  mass term.  

The induced Chern-Simons term can be seen in terms of diagrammatic perturbation theory or using the Landau level arguments along the lines of \cite{redlichL, redlichD, dunne-rev}. The latter approach is closer to the Hamiltonian
formulation and we review it here in terms of our gauge-invariant variables.
Since the putative CS term in the effective action has the lowest number of derivatives, it suffices to consider constant magnetic fields. For the Abelian case, we can take
$A_i = \half B \epsilon_{ij} \, x_j$, $\epsilon_{12} =1$, and the Schr\"odinger equation becomes
\be
\left(\begin{array}{cc}
m & -2iP_-\\
2iP_+ & -m
\end{array}\right) \left(\begin{array}{c}\psi_1 \\ \psi_2\end{array}\right) = E\left(\begin{array}{c}\psi_1 \\ \psi_2\end{array}\right)\label{ev}
\ee
where
\be
P_+ = -i\del -\frac{i}{4} B\bar z, \hskip .3in
P_- = -i\bar\del + \frac{i}{4}Bz
\label{P-def}
\ee
Here $\del = \frac{1}{2}(\del_i + i\del_2)$ and $z = x_i -ix_2$. In the complex representation $A = \frac{i}{2}B\bar z, \bar A = -\frac{i}{2}Bz$. Since $[P_+, P_- ] = \half B$, we can easily diagonalize $P_- P_+$
and $P_+ P_-$ as
\be
P_-P_+ = \frac{B}{2}\, n \hspace{1cm} P_+P_- = \frac{B}{2}\, (n+1)
\ee 
where $n$ is either zero or a positive integer. The eigenvalue equation (\ref{ev}) gives
$E^2 = m^2 + 4 P_- P_+$, so that the states with the lowest value
of $\vert E\vert$ have $E = \pm m$. For 
$m >0$ and $E = +m$, (\ref{ev}) reduces to 
\be
P_-\psi_2 = 0, \hspace{.2cm} 2i P_+\psi_1 = 2m\psi_2 
\label{zeromode}
\ee
From the expressions (\ref{P-def}), we see that only $P_+$ can have normalizable zero modes, 
hence the solution to (\ref{zeromode})  is given by 
\be
\Psi = \left(\begin{array}{c}\psi_1 \\ 0\end{array}\right), \hspace{1cm} P_+\psi_1 = 0 
\label{cons}
\ee
Consider now the
charge density defined as
 $J_0 = \frac{1}{2}[\Psi^\dagger, \Psi]$. 
This definition is required by the property that $J_0$ should be odd under charge conjugation.
The zero-mode  in (\ref{cons}) will contribute to the vacuum expectation value
of $J_0$. 
This can be seen as follows. The mode expansions of fermion fields are
\begin{eqnarray}
\Psi = a \Psi_0 + \sum_n a_n\Psi^+_n + \sum_nb^\dagger_n\Psi^-_n\\
\bar \Psi = a^\dagger \bar \Psi_0 + \sum_n a_n^\dagger \bar \Psi ^+_n + \sum_n b_n\bar \Psi ^-_n
\end{eqnarray}
The charge density can then be evaluated as
\be
J_0 = \sum_n(a^\dagger_na_n - b^\dagger_n b_n) + \frac{1}{2}(a^\dagger a - a a^\dagger) 
\ee
The last term on the right is the zero mode contribution. 
The vacuum is a state where all the negative energy states are filled; in the present case, since the
zero mode has energy $+m$, the vacuum state $\vert \Omega\ra$ is the Fock vacuum 
$\vert 0\ra$ defined by
$a \, \vert 0 \ra = a_n \vert 0 \ra = b_n \vert 0 \ra = 0$. Thus
the expectation value of the charge density for this state is given by
\be
\langle 0| J_0|0\rangle = \frac{e^2B}{2\pi}\langle 0|\frac{1}{2} (a^\dagger a - a a^\dagger) |0\rangle = -\frac{e^2B}{4\pi}\label{j01}
\ee
where we have taken account of the fact that there is also a degeneracy of $\frac{eB}{2\pi}$
for the Landau levels.
Repeating the analysis for $m<0$ and $E = -|m|$, we see that the zero mode solution given above  is again a normalizable zero mode, which, however,  must now be regarded as part of the negative energy states. Thus the vacuum $|\Omega\rangle$ is not the Fock vacuum $\vert 0\ra$, but must be taken to be $|\Omega\rangle = a^\dagger |0\rangle$.
As a result,
\be
\langle \Omega | J_0|\Omega \rangle =   +\frac{e^2B}{4\pi}\label{j02}
\ee
Combining (\ref{j01},\ref{j02}), 
\be
\langle J_0\rangle = -\frac{m}{|m|}\frac{e^2B}{4\pi} =  -\frac{m}{|m|}\frac{e^2}{8\pi} \frac{\delta}{\delta A_0}\int \epsilon^{\mu \nu \rho}A_\mu \partial_\nu A_\rho
\ee
This shows that the contribution to the effective action upon integrating out the fermions
and including both electric and magnetic fields
is
\be
S_{eff} = -\frac{m}{|m|} \frac{e^2}{8\pi}\int A\wedge dA\label{scs}
\ee
Thus the induced level number $k$ in the Abelian case is $-m/|m|$,
or $k = -\frac{m}{|m|} \, N_f$ for 
Abelian theories with a $U(N_f)$ flavor symmetry.

This result agrees with the induced Chern-Simons term computed in a path integral picture,
as in \cite{dunne-rev}. This analysis also generalizes readily to nonabelian fundamental fermions. One can take the gauge field $A^a$ to be nonzero only in a fixed direction of the color space.  For an $SU(N)$ gauge group, for example, we can take $A^1_i= \cdots   = A^{N^2-2}_i = 0$, $A^{N^1-1}_i = -\frac{1}{2}\epsilon_{ij}x_j B^{N^2-1}$, where $B^{N^2-1}$ is the constant nonvanishing component of the magnetic field. Generalizing the Abelian analysis to the $N$-plet of fermions, we
obtain
\be
\langle J^{ij}\rangle = \langle [\psi^{\dagger i}, \psi^j]\rangle =  -\frac{m}{|m|}\frac{e^2B^{N^2-1}}{4\pi}\left(t^{N^2 - 1}\right)^{ij}
\ee
Here the $SU(N)$ generators $t^i$ are normalized as $\tr (t^at^b) = \frac{1}{2}\delta^{ab}$. The nonvanishing component of the current $J^{N^2-1} = J^{ij} \left(t^{N^2 - 1}\right)^{ji}$ is now related to the variation of the Chern-Simons term as
\be
\langle J^{N^2-1}\rangle =  -c_F\frac{m}{|m|}\frac{e^2}{4\pi} \frac{\delta}{\delta A^{N^2-1}_0}\int d^3x \,\epsilon^{\mu \nu \alpha}~\Tr \left(A_\mu\partial_\nu A_\alpha + \frac{2}{3}A_\mu A_\nu A_\alpha\right)
\ee 
$c_F = \frac{1}{2}$ is the normalization for the $SU(N)$ generators and $A_\mu$ are in the anti-hermitian basis $A_\mu = -i t^a \, A^a_\mu$.

The basic point of the analysis given above is  to emphasize that the induced Chern-Simons term
can be understood in terms of fermionic zero modes, i.e., the lowest Landau level in a constant magnetic field. We did not include a `bare' Chern-Simons term, but if we do so, say, with level number 
$k$, the fermion contribution amounts to a renormalization
$k \rightarrow k + c_F (m / \vert m\vert ) $. This renormalization is equivalent to a one-loop
contribution is diagrammatic perturbation theory, and it is, of course, known that there are no further
corrections from higher loops.

The existence of zero modes obviously holds in terms of our gauge-invariant variables as well,
as it is guaranteed by an index theorem. We can also show this explicitly
 for each of our choices for the gauge invariant variables, see appendix B.
 So thinking entirely in terms of the gauge-invariant variables, how does one put together
 the CS term and the Jacobian we found earlier? If we set up the Hamiltonian in terms of the gauge-invariant variables, we can, in principle, obtain the path integral using $e^{-i \H t}$ 
 and carrying out the time-slicing in the usual way.
 The CS action can then be captured again via diagrammatic perturbation theory, but
 now in terms of the gauge-invariant variables. However, recall that the transformation
 to the gauge-invariant variables is like a complex gauge transformation
 for the $A_i$ and that the CS term can generate boundary terms upon carrying out such transformations. While spatial boundary terms may be neglected, the temporal boundary terms
 remain. These lead to $S_{WZW}(H)$ as may be seen in the Hamiltonian analysis of
 the CS action in \cite{bos-nair}.
 In other words, the Jacobian in the measure of integration should be viewed as
 what is needed to complete the CS term which would be obtained from the diagrammatic analysis of
 $e^{- i \H t}$ (say, via path integrals) using $\H$ in terms of the gauge-invariant variables.
 The argument presented in \cite{AA-VPN} showed how the CS term via the Knizhnik-Zamolodchikov
 equations led to the form of the gauge-invariant measure. The present argument completes the circle of starting with the Hamiltonian and recovering the CS action.
 
For the sake of completeness of this argument, we must consider the zero modes of the Landau problem for the adjoint fermions. Therefore, a brief discussion of the adjoint case is 
appropriate at this point.
 
\subsection{Fermions in the adjoint representation}

 The computation of the fermionic Jacobian in section II  showed that contributions of adjoint fermions can, in principle, be nontrivial. Choices I and II were shown to contribute $ e^{\pm (c_A S_{WZW}[H])}$ for each Majorana species. The argument of the exponential doubles when these choices are employed for Dirac fermions. 
To illustrate the point under discussion,
it is sufficient to consider the case of the gauge group being $SU(2)$.
We take the structure constants to be given by $\e^{ijk}; \e^{123} = +1$. We can choose the constant
magnetic field to be along the third direction in color space and 
fix a gauge such that
$A^1 = \bar A^1 = A^2 = \bar A^2 = 0$ and $A^3_i =  -\frac{1}{2}\epsilon_{ij}x_j B$. 
The covariant derivative $(D\Psi)^a = \partial \Psi^a + \e^{abc}A^b\Psi^c.$
We start with a Dirac fermion $\Psi$ with upper component
$u^a$ and lower component $v^a$.  We will not display the color index for the 
gauge field from here on, unless it is needed for clarity.
The Dirac equation becomes
\be
\left(\begin{array}{cc}
m & -2\bar D\\
2D & -m
\end{array}\right) \left(\begin{array}{c} u\\ v\end{array}\right) = E\left(\begin{array}{c}u\\ v\end{array}\right)\label{maj}
\ee
In writing the above form of the equation, we have assumed that the mass for the Dirac fermion is $\int \bar \Psi \Psi$; it is a parity violating mass term.
Written out  in terms of the color components, these equations become
\begin{eqnarray}
-2\bar \Delta_{\pm} v_{\pm} &=& (E-m)u_{\pm} \nonumber\\
+2 \Delta_\pm u_\pm &=& (E+m)v_\pm 
\label{Dirac-Maj}
\end{eqnarray}
where $u_\pm = (u^1 \pm i u^2)/\sqrt{2}$, $v_\pm = (v^1 \pm i v^2)/\sqrt{2}$ and
\begin{align}
&\bar \Delta_+  = \bar \del + i\bar A = \bar \del + \frac{1}{4}zB, 
&\Delta_+  = \del + i A = \del - \frac{1}{4}\bar z B\nonumber\\
&\bar \Delta_-  = \bar \del - i\bar A = \bar \del - \frac{1}{4}zB,
&\Delta_-  = \del - i A = \del + \frac{1}{4}\bar z B\label{Delta}
\end{align}
Note that only $\bar \Delta_+ $ and $\Delta_-$ can have a nontrivial normalizable zero mode given by $\omega  = \exp\left({-\frac{1}{4}z\bar z B}\right)$.
It is useful to note that
while $[ \Delta_+, \bar \Delta_+] \neq 0$ and
$[ \Delta_-, \bar\Delta_- ] \neq 0$, 
the pair $( \Delta_+, \bar\Delta_+)$ commutes with the pair
$( \Delta_-, \bar\Delta_- )$; thus one pair acts as the magnetic translations for the other pair.

Equations (\ref{Dirac-Maj}) imply the existence of zero mode solutions. For example, when $E = +m$, $m\ge 0$, we have the solution $v_\pm = 0, u_+ = 0, u_- = \exp\left( - {1\over 4}Bz\bar z\right)$. It is straightforward to apply the chain of reasoning employed in the analysis of the lowest
Landau level for fundamental fermions to this case and show that the nontrivial zero mode can account completely for the induced Chern-Simons term for the parity violating adjoint Dirac fermion. Using the solution obtained above one can compute the induced current in the gauge employed above.
\be
J^3 = -\frac{i}{2}\langle[\bar \Psi^i, \Psi^j]\rangle \e^{3ij}= -c_A\frac{m}{|m|}\frac {e^2B^3}{4\pi}
\ee
The gauge invariant form of the current is then seen to be given by
\be
\langle J^{3}\rangle =  -c_A\frac{m}{|m|}\frac{e^2}{4\pi} \frac{\delta}{\delta A^{3}_0}\int d^3x ~\Tr \left[\left(A_\mu\partial_\nu A_\alpha + \frac{2}{3}A_\mu A_\nu A_\alpha\right)\epsilon^{\mu \nu \alpha}\right].
\ee 
The induced Chern-Simons term correctly accounts for the renormalization of a bare $k$ to $k - c_A\frac{m}{|m|}$, in agreement with standard results from covariant perturbation theory. 
We can conclude that choices I and II are appropriate for Dirac fermions with a parity-violating mass term.

Turning now to the Majorana case, we first note that the Landau level argument
for the CS term relies on the definition of $J_0$ which makes it odd under
charge conjugation. Thus we should not expect to use the zero mode analysis 
for the Majorana case. In fact, if we consider a Majorana fermion, we need 
$v = u^*$ in (\ref{maj}).
The eigenvalue equation is thus
\be
\left(\begin{array}{cc}
m & -2\bar D\\
+2D & -m
\end{array}\right) \left(\begin{array}{c}u \\ u^*\end{array}\right) = E\left(\begin{array}{c}u \\ u^*\end{array}\right)\label{maj2}
\ee
Defining the complex combinations $ \phi = (u^{1} - i u^{2})/\sqrt{2}$, $\chi =  (u^{1} + i u^{2})/\sqrt{2}$,
these equations become
\begin{eqnarray}
-2\bar\Delta_+\phi^* = (E-m)\chi, \hspace{1cm}+2\Delta_+\chi = (E+m)\phi^*\nonumber \\
-2\bar \Delta_- \chi^* = (E-m)\phi, \hspace{1cm} +2\Delta_-\phi = (E+m)\chi^*\label{maj1}
\end{eqnarray}
The spectrum of the products of $\Delta$'s  is readily found to be
\be
\bar\Delta_+\Delta_+ = \Delta_-\bar\Delta_- = {B \over 2} \,(n+1), \hspace{1cm} \Delta_+\bar \Delta_+ = 
\bar\Delta_- \Delta_-  = {B \over 2} \,n
\ee
It is easy to see that the lowest eigenvalue for (\ref{maj1}) corresponds to $E = \pm m$. When $m>0$, and $E = +m$, (\ref{maj1}) leads to
\be
\bar\Delta_-\chi^* = \bar \Delta_+\phi^* = 0, \hspace{1cm} \Delta_-\phi = m\chi^*, 
\hspace{1cm}\Delta_+\chi = m \phi^*
\ee
Since $\bar\Delta_-$ cannot have a normalizable zero mode,
the first equation implies $\chi ^* = 0 = \chi$. This, when used in the last equation gives $\phi^* = 0 = \phi$.
For $m>0$ and $E = -|m|$, (\ref{maj1}) leads to
 \be
 \Delta_-\phi = \Delta_+\chi = 0,\hspace{1cm} \bar\Delta_-\chi^* = m\phi, \hspace{1cm}\bar\Delta_+\phi^* = m\chi
 \ee
Again the first equation implies $\chi = 0 = \chi^*$, which, in turn gives $\phi  = 0 = \phi^*$ through the second equation. 
Thus, no nontrivial zero-modes exist for the $SU(2)$ Majorana fermions. (This analysis can be easily generated to general a $SU(N)$ gauge group.)

How should we consider this result in relation to the nontrivial measure for Majorana fermions?
The point is that the fermionic determinant in the covariant path integral for Majorana
fermions must be defined by an appropriate square root of the Dirac determinant.
There is no obstruction to regarding the Dirac fermion as a pair of Majorana fermions.
In fact, for the
Dirac fermion, we can write
$\Psi = (\chi_1 - i \, \chi_2 )/\sqrt{2}$, where the $\chi_i$ are Majorana fields.
This translates to the level of the gauge-invariant variables as well.
Let $G(\chi)_i$ refer to the gauge invariant version of $\chi$ for either
choice I or II. If $\chi_i$ have the same parity violating mass, then
\be
\int {\bar \Psi} \, \H \, \Psi = \sum_i \int {\bar \chi}_i \, \H \, \chi_i
\label{maj3}
\ee
where $\H$ is the single particle Hamiltonian. This is the usual split of the Dirac fermion
into two Majorana modes. However, it is also straightforward to see that
\be
\sum_i \int G({\bar\chi}_i) \, \H_J \, G (\chi_i ) =
\int G({\bar \Psi} )\, \H_J \, G(\Psi) 
\label{maj4}
\ee
where $\H_J$ refers to the single particle Hamiltonian expressed in the gauge-invariant
variables.
Thus, when we have two Majorana species - as we did in the ${\cal N}= 2$ supersymmetric
case - the Hamiltonian dynamics is completely equivalent to having a single Dirac species.
This shows that the choice of gauge invariant variables for parity-violating Dirac
fermions must be choices I or II, which of course, have nontrivial Jacobians. Thus
the $A_0 B$ term generated by the zero modes of the Dirac fermions 
give the completion of the Jacobian contribution, as before. The 
induced CS term for the Majorana case will follow from the Dirac case by taking the
proper square root.
\subsection{Selection rules for fermionic variables}

We are finally in a position to summarize our results as a set of rules on how the gauge-invariant
versions of the fermion fields are to be chosen.

As far as fermions with parity violating mass-terms are concerned, one has to fix a parity frame in which to formulate the theory since parity is not a symmetry. Let us choose the parity frame corresponding to (\ref{QCDF}). 
Putting together the computation of the fermionic Jacobians and the induced Chern-Simons terms we see that the only consistent choice of gauge-invariant variables (for each fermionic species corresponding to a parity violating mass term) corresponds to choosing Choice II when $m>0$ and Choice I for $m<0$. It is understood that  choices I and II refer to (\ref{19}) when the fermions are in the fundamental representation. In a theory with $d$ species of parity odd Dirac fermions with masses $\mu_i$ (in representation $R_i$),  $m$ species of adjoint Majorana  fermions with masses $M_i$ and a Chern-Simons terms with bare level number $k_{CS}$,  the coefficient of $S_{WZW}$ would be renormalized as
\be
\tilde k = 2c_A \rightarrow \tilde k + k', \hspace{.2cm} k' =  + k_{CS} + 2\sum_{i=1}^d \frac{\mu_i}{|\mu_i|}c_{R_i} + \sum_{i=1}^m  \frac{M_i}{|M_i|}c_{A}\label{levelr}
\ee
The expression on the $r.h.s$ follows from the prescription of selecting variables outlined above and is the unique choice that is consistent with (\ref{consistency}). Of course, it is not a given that the expression on the $r.h.s$ is positive definite. However, one can always find a parity frame where it is, and that is the frame that one must use to define the theory.

Fermions with parity-conserving mass terms do not induce Chern-Simons couplings. One could generate a parity-symmetric theory by having an equal number of parity-violating Dirac or Majorana fermions with masses of opposite signs. In that case, the above selection principle can be applied directly. However, in the case of an even number of fermionic species one can also define parity conserving mass-terms that mix the different species. In such a scenario, Choices I and II must be ruled out. Of the two remaining choices, one does not have a clear principle to pick one. However, we note that  Choice III leads to a fermionic Hamiltonian involving only $J$ (and no $\bar J$). Since $J$ is realized straightforwardly in its position space representation in the gluonic Hilbert space, choosing 
III is computationally advantageous when there are no parity-related obstructions to making that choice. 

\section{Discussion}

In this section we comment on how the fermionic contribution 
(or lack thereof) to $d\mu(\mathcal{C})$ can impact on questions of screening and confinement.
If the fermions are taken to be very heavy (compared to the mass scale $e^2$ given by the coupling constant), the vacuum wave function will factorize as $\Phi = \Phi_g \otimes \Phi_f$, where
$\Phi_f$ is the fermionic vacuum and $\Phi_g$ is determined by the gluonic part of
the Hamiltonian. The kinetic operator for the latter has ${\tilde k} + k'$ due to the fermionic contribution $k'$. The gluonic Hamiltonian then has a form similar to what was obtained for the YMCS theory and
we can exploit this similarity to write the leading strong coupling part of $\Phi_g$ as
\cite{YMCS}
\be
\Phi_g = e^{-\frac{1}{8\,g^2} F^a_{ij}F^{aij}}, \hspace{.2cm} g^2 = \frac{e^4(\tilde k + k')}{4\pi}\label{vac}
\ee
Vacuum expectation values of observables, $\mathcal{O}$, can be expressed as
\be
\langle \mathcal{O} \rangle = \int d\mu[H] e^{(\tilde k + k')S_{WZW}(H)}e^{-\frac{1}{4g^2} F^a_{ij}F^{aij}}\mathcal{O} \label{wilson}
\ee
For $k' =0$, this integral can be evaluated by relating it to the Euclidean
two-dimensional Yang-Mills theory.
For $k' \neq 0$, 
 the computation of the integral in (\ref{wilson}) is completely equivalent to that of the YMCS system with the Chern-Simons level number given by $k'$. (It can also be viewed as the computation of
 a correlation function in Euclidean $QCD_{2}$ with $k'$ flavors of fundamental $SU(N)$ fermions. )
 As shown in \cite{YMCS}, taking ${\cal O}$ as a Wilson loop,
 for large spatial loops enclosing an area $\mathcal{A}$, the expectation value
 falls off as $e^{-w}$, where $w/\mathcal{A} \rightarrow 0$ as $\mathcal{A} \rightarrow \infty$.
In other words, we do not have an area law; a nonzero contribution from the fermions to the measure results in screening, as opposed to confinement of color charges.  
This statement is independent of the representation the fermions may be in, and continues to hold even when $k_{CS} = 0$. Although we can argue for this statement
only in the leading strong coupling approximation for heavy fermions,
it is reasonable to expect the screening behavior to persist when dynamical contributions from the fermionic wave function are incorporated.

For $k' = 0$, {\it a priori} we would conclude that the Wilson loops do display an
area law, with $\la W_R \ra \sim \exp ( - \sigma_R \, {\cal A} )$ 
for a Wilson loop $W_R$ in the representation $R$, with
$\sigma_R = e^4 c_A c_R / 4 \pi$ as for the pure Yang-Mills case.
However, there is a subtlety.

As shown in \cite{String-Breaking}, adjoint matter fields (more generally matter fields of vanishing $N$-ality) can lead to the formation of `glue-lump' states (bound states of the fermions and $J$ fields) which will be energetically favored for large quark separation. Thus we may expect the 
 potential between adjoint heavy quarks to rise linearly, $V(r) \sim \sigma_A |r|$ for some 
 intermediate range of $r$, but for large enough values of the separation $r$, the adjoint string is expected to break into glue-lump states, leading to a flat inter-quark potential.
(Though \cite{String-Breaking} focussed on scalar matter, the same mechanism would apply to adjoint fermions as well.)
The  numerical value of $r$ for the string to break requires redoing the analysis of  
\cite{String-Breaking} for fermionic matter. But qualitatively, we expect screening behavior.

For fermions in the fundamental representation, which give $k' =0$, we recover the area law for Wilson loops. As long as the fermions are heavy, the area law can be expected to hold, signaling confinement as expected. However, when the fermions are dynamical, there is  the potential for screening via the pair production of fermions. Understanding how confinement can be manifest with dynamical fundamental fermions would require a better understanding of the vacuum wave functional, going beyond the heavy quark approximation employed in this discussion.

We have presented a qualitative picture of screening and confinement based on a gauge invariant formulation of $QCD_{2+1}$, focusing on the fermionic contribution to
the volume measure for the gauge-invariant 
configuration space.
While our statements about confinement and string tension are exact in the leading strong coupling and heavy quark approximation, this does open up
several avenues for further exploration.
Corrections to the leading heavy quark approximation are obviously important.
It should be possible to formulate this along the lines of corrections to the
vacuum-wave functional (\ref{vac}) for pure Yang-Mills theory \cite{correction}. 
The inclusion of matter fields in arbitrary representations of the gauge group would be
another interesting direction.

In an interesting series of papers \cite{unsal}, gauge theories on $R^{1,3}\times S^1$ have been analyzed. It has been argued in these papers that several nonperturbative aspects of four-dimensional gauge theories can be studied by taking $S^1$ to be small. It would be extremely interesting to apply the present formalism to the class of theories studied in \cite{unsal} and study their  properties from  the inherently three-dimensional point of view developed in this paper.

Finally, a rather rich landscape of supersymmetric Yang-Mills and Chern-Simons theories (beyond the class of models studied in \cite{AA-VPN}) remains to be explored. The methods developed here can be applied directly to supersymmetric theories, which is something we hope to pursue elsewhere.

\bigskip
This work was supported in part by the U.S.\ National
Science Foundation grant PHY-1213380 and by a PSC-CUNY award.
\section*{Appendix A}\label{app1}
The Jacobian can be computed using the standard techniques of chiral anomaly computation in two dimensions. Focusing on the part that involves $\gamma^5$, 
 \be
 \det C = \mp \Tr\int (\Re \theta^aT^a\gamma^5) + \cdots  = \mp \lim_{M\rightarrow \infty} \Tr\int (\Re\theta^aT^a\gamma^5 e^{-{(D.\gamma})^2/M^2}) + \cdots
 \label{appA1}
 \ee
 The ellipsis denote terms that are higher order in $\theta$. $D.\gamma$ is the Hermitian two dimensional Euclidean Dirac operator $D.\gamma = i\gamma^i\partial_i + T^aA^a_i\gamma^i$. A straightforward computation shows that
 \be
 \det C = \mp\int \frac{c_A}{2\pi} F^a_{12}(\Re \theta )^a \cdots \label{jack1}
 \ee
 This answer is still in the infinitesimal form. To get the answer for finite value of $\Re \theta$ we note that 
 \be
 \frac{\delta \det C}{\delta \Re \theta^a} = \mp \frac{c_A}{2\pi} F^a_{12} = \frac{\delta e^{\pm c_A S_{WZW}(H)}}{\delta \Re \theta^a}
 \ee
 which we can integrate to obtain the result
 \be
 \det C =  e^{\pm (c_A S_{WZW}[H])}
 \ee
 We note that the final answer only depends on $\Re \theta$ which is the gauge-invariant dynamical degree of freedom contained in the gauge fields. The imaginary part of $\theta $ encodes the unphysical degree of freedom associated with gauge transformations.

For the charge conjugate representation we have
 \be
 \det \tilde C =  \det (I + i\Im \theta^a  (T^t)^ a \pm \Re \theta^a (T^t) ^a\g^5) =  \pm \lim_{M\rightarrow \infty}\Tr\int (\Re\theta^a(T^t)^a\g^5 e^{-{(\tilde D.\gamma})^2/M^2})+ \cdots\label{jack3}
 \ee
Here $\tilde D$ is the hermitian covariant derivative acting on fields transforming in the charge conjugate representation. The precise form of $\tilde D$ can be derived as a follows. On states $\Psi $ that transform as $\Psi \rightarrow U \Psi$ (where $U = e^{iT.\vf}$ is an $SU(N)$ group element), $D_i =  i\partial_i + A$ is the correct covariant derivative with $A_i = A^a  T^a$ transforming as $A_i \rightarrow U\,A_i\,U^{-1} -i\partial_iU\,U^{-1}$. 

The charge conjugate fields ${\tilde \Psi}$ transform as $\Psi^*$ does, i.e.,
${\tilde \Psi} \rightarrow U^*\, {\tilde \Psi}$.
Since $U^* = (U^\dagger)^t = (U^{-1})^t =  e^{- iT^T.\vf}$, we see that
we need $-A_i^t$ in place of $A_i$ in the covariant derivative. In other words,
for the conjugate representation, ${\tilde D}_i  {\tilde \Psi} = ( i \partial_i - A_i^t ){\tilde \Psi}$, with
 \be
 {\tilde D}_i {\tilde \Psi} \rightarrow ({ \tilde U}^{-1} )^t\, ({\tilde D}_i {\tilde \Psi})
 \ee
The needed transformation of $-A_i^t$, namely, $-A_i^t \rightarrow
(U^{-1})^t \, (-A_i^t) \, U^t + i (U^{-1})^t\, \partial_i U^t$ is then compatible with
$A_i \rightarrow U\,A_i\,U^{-1} -i\partial_iU\,U^{-1}$.
Since the $F_{12}$ term in (\ref{jack1}) arises from $[D_1, D_2]$,  the corresponding term in (\ref{jack3}) - involving  $[\tilde D_1, \tilde D_2]$ will have a negative sign relative to (\ref{jack1}). This is due to the sign of  $\tilde A$ in $\tilde D$ being the opposite of $A$ in $D$.  Thus
 \be
 \det \tilde C = \mp\int \frac{c_R}{2\pi} F^a_{12}(\Re \theta )^a \cdots \label{jack4}
 \ee
 where $R$ refers to the representation of interest. Thus, quite generally, we have:
 \be
 \det C = \det \tilde C =  e^{\pm (c_R S_{WZW}[H])}
 \ee
 The detrrminant is the same for a given representation and its conjugate. The coefficient
$c_R = \half $ for the fundamental representation of $SU(N)$ for $N\geq 2$. In the Abelian case, $c_R = 1$. In the adjoint representation, $c_R = c_A = N$.
 
\section*{Appendix B}\label{app2} 

{\bf Choice I}
\vskip .1in\noindent
The gauge-invariant form of the  fermionic Hamiltonian is
\be
H = \frac{1}{2} \int \Lambda^\dagger\left(\begin{array}{cc}
Hm & -2\bar\del \\
2\del & -H^{-1}m
\end{array}\right)\Lambda
\ee
Naively one might try to solve the eigenvalue equation 
\be
\left(\begin{array}{cc}
Hm & -2\bar\del \\
2\del & -H^{-1}m
\end{array}\right)\Lambda = E \Lambda
\ee
It is easily seen that when $\del $ or $\bar \del$ is set to zero the only solution is 
 $\chi_1 = \chi_2 =0$.
Thus the Landau levels along with the nontrivial zero modes appear to have disappeared in this 
gauge-invariant parametrization. 
However, this is not quite the correct thing to do. If we want to diagonalize the Hamiltonian, we should instead solve 
\be
\left(\begin{array}{cc}
Hm & -2\bar\del\\
2\del & -H^{-1}m
\end{array}\right)\Lambda = E  \left(\begin{array}{c}H\chi_1\\ H^{-1}\chi_2 \end{array}\right)
\ee
This equation is nothing but (\ref{ev}) written in terms of the gauge-invariant variables. Thus its solutions are the same as the solutions of the original equation (\ref{ev}), but re-expressed in terms of $\chi_i$. The diagonalized Hamiltonian now takes on the form
\be
H = \int \sum_n(\chi_1^* H \chi_1 + \chi_2^* H^{-1} \chi_2)_nE_n
\ee
The subscripts refer to the energy eigenvalue. Clearly the zero modes that were present in the original set up are now present in the gauge-invariant language with the same energy eigenvalues.
\vskip .1in\noindent
{\bf Choice II}
\vskip .1in\noindent
In this case, the fermionic Hamiltonian is
\be
H = \frac{i}{2}\int \bar \Psi (\gamma^iD_i - m)\Psi = \frac{1}{2} \int \Lambda^\dagger\left(\begin{array}{cc}
H^{-1}m & -2i\Pi_-\\
2i\Pi_+ & -Hm
\end{array}\right)\Lambda
\ee
\be
\Pi_+ = -i\mathcal{D}_+ = -i(\del - \del HH^{-1})\hskip .2in \Pi_- = -i\bar{\mathcal{D}}_{-} = -i(\bar \del + H^{-1}\bar \del H)
\ee
The equation we now need to solve for diagonalizing the Hamiltonian
 in the gauge-invariant variables is
\be
\left(\begin{array}{cc}
mH^{-1} & -2i\Pi_-\\
2i\Pi_+ & -mH
\end{array}\right) \left(\begin{array}{c}\chi_1 \\ \chi_2\end{array}\right) = E\left(\begin{array}{c}H^{-1} \chi_1 \\ H\chi_2\end{array}\right)\label{dirac2}
\ee
Once again this just (\ref{ev}). 
To find the lowest lying states, it is useful to  note that in a constant magnetic field, we have  $M = e^{-\frac{1}{4} z\bar zB}$, which is compatible with the gauge choice $A_i = -\frac{1}{2}\epsilon_{ij}x_j B$ as well as the parameterization $A = -\del MM^{-1}$. In this parameterization, $\Pi_\pm$ are the same as $P_\pm$ with $B \rightarrow 2B$.The spectrum of the bilinears is readily found to be $\Pi_+\Pi_- = B(n+1)$, $\Pi_-\Pi_+  = nB$.
It is straightforward to see that the solutions of (\ref{dirac2}) are the same as those (\ref{ev}) with the diagonalized Hamilatonian now being
\be
H = \int \sum_n(\chi_1^* H^{-1} \chi_1 + \chi_2^* H \chi_2)_nE_n
\ee
The eigenvalues, including the zero modes, remain unchanged.
\vskip .1in\noindent
{\bf Choices III \& IV}
\vskip .1in\noindent
The Hamiltonians are 
\be
H = \frac{1}{2} \int \Lambda^\dagger H^{-1}\left(\begin{array}{cc}
m & -2\bar \del \\
2\mathcal{D}_+ & -m
\end{array}\right)\Lambda
\ee
and 
\be
H = \frac{1}{2} \int \Lambda^\dagger H\left(\begin{array}{cc}
m & -2\bar{\mathcal{D}}_{-}\\
2\del & -m
\end{array}\right)\Lambda
\ee
respectively. It is easy to see that the eigenvalue equations for these choices are the same as (\ref{ev}) but in the gauges $A = iB\bar z, \bar A = 0$ and $A = 0, \bar A = -iBz$ respectively. Thus the zero modes along with the degeneracies associated with the Landau levels are not removed by these last two choices of gauge invariant variables either.



\begin{thebibliography}{99}
\bibitem{KKN1} 
  D.~Karabali and V.~P.~Nair,
  Nucl.\ Phys.\ B {\bf 464}, 135 (1996)
  [hep-th/9510157],
  D.~Karabali and V.~P.~Nair,
  Phys.\ Lett.\ B {\bf 379}, 141 (1996)
  [hep-th/9602155],
  D.~Karabali, C.~j.~Kim and V.~P.~Nair,
  Nucl.\ Phys.\ B {\bf 524}, 661 (1998)
  [hep-th/9705087]. 


\bibitem{AA-VPN} 
  A.~Agarwal and V.~P.~Nair,
  Phys.\ Rev.\ D {\bf 85}, 085011 (2012)
  [arXiv:1201.6609 [hep-th]].

\bibitem{dunne-rev} 
  G.~V.~Dunne,
  hep-th/9902115.
  
  \bibitem{String-Breaking} 
  A.~Agarwal, D.~Karabali and V.~P.~Nair,
  Nucl.\ Phys.\ B {\bf 790}, 216 (2008)
  [arXiv:0705.0394 [hep-th]].

\bibitem{YMCS} 
  D.~Karabali, C.~-j.~Kim and V.~P.~Nair,
  Nucl.\ Phys.\ B {\bf 566}, 331 (2000)
  [hep-th/9907078].

\bibitem{robust} 
  D.~Karabali and V.~P.~Nair,
  Phys.\ Rev.\ D {\bf 77}, 025014 (2008)
  [arXiv:0705.2898 [hep-th]].
  
 \bibitem{YMCS-lee} 
  H.~C.~Kao, K.~M.~Lee and T.~Lee,
  Phys.\ Lett.\ B {\bf 373}, 94 (1996)
  [hep-th/9506170].
  
 \bibitem{bos-nair} 
  M.~Bos and V.~P.~Nair,
  Int.\ J.\ Mod.\ Phys.\ A {\bf 5}, 959 (1990). 
  
\bibitem{rob-rao} 
  R.~D.~Pisarski and S.~Rao,
  Phys.\ Rev.\ D {\bf 32}, 2081 (1985).  

\bibitem{wittencs} 
  E.~Witten,
  Commun.\ Math.\ Phys.\  {\bf 117}, 353 (1988).  
  
\bibitem{jackiw} 
  R.~Floreanini and R.~Jackiw,
  Phys.\ Rev.\ D {\bf 37}, 2206 (1988). 

\bibitem{wipf} 
  C.~Kiefer and A.~Wipf,
  Annals Phys.\  {\bf 236}, 241 (1994)
  [hep-th/9306161].
  
 \bibitem{mansfield} 
  P.~Mansfield and D.~Nolland,
  Int.\ J.\ Mod.\ Phys.\ A {\bf 15}, 429 (2000)
  [hep-th/9907159].

\bibitem{redlichL} 
  A.~N.~Redlich,
  Phys.\ Rev.\ Lett.\  {\bf 52}, 18 (1984).

\bibitem{redlichD} 
  A.~N.~Redlich,
  Phys.\ Rev.\ D {\bf 29}, 2366 (1984).
  
 \bibitem{an-book} 
  R.~A.~Bertlmann,
  Oxford, UK: Clarendon (1996) 566 p. (International series of monographs on physics: 91) 
  
\bibitem{correction} 
  D.~Karabali, V.~P.~Nair and A.~Yelnikov,
  Nucl.\ Phys.\ B {\bf 824}, 387 (2010)
  [arXiv:0906.0783 [hep-th]].
\bibitem{unsal} 
  M.~Unsal and L.~G.~Yaffe,
  Phys.\ Rev.\ D {\bf 78}, 065035 (2008)
  [arXiv:0803.0344 [hep-th]],
  E.~Poppitz and M.~Unsal,
  JHEP {\bf 1001}, 098 (2010)
  [arXiv:0911.0358 [hep-th]],
M.~Unsal and L.~G.~Yaffe,
  JHEP {\bf 1008}, 030 (2010)
  [arXiv:1006.2101 [hep-th]],    
  M.~Shifman and M.~Unsal,
  Phys.\ Lett.\ B {\bf 681}, 491 (2009)
  [arXiv:0901.3743 [hep-th]],
  A.~Armoni, D.~Dorigoni and G.~Veneziano,
  JHEP {\bf 1110}, 086 (2011)
  [arXiv:1108.6196 [hep-th]]. 
\end{thebibliography}
\end{document}